\documentclass[12pt]{amsart}

\usepackage[dvipdfm]{graphicx}
\usepackage[dvips]{geometry}
\usepackage{color}

\usepackage{amssymb}
\newcommand{\be}{\begin{equation}}
\newcommand{\ee}{\end{equation}}
\newcommand{\cE}{{\mathcal E}}
\newcommand{\CC}{{\mathbb C}}
\newcommand{\CI}{{\mathcal C}^\infty }
\newcommand{\CIc}{{\mathcal C}^\infty_{\rm{c}} }

\newcommand{\RR}{{\mathbb R}}

\newcommand{\defeq}{\stackrel{\rm{def}}{=}}

\newcommand{\vol}{\operatorname{vol}}

\newcommand{\supp}{\operatorname{supp}}

\newcommand{\rest}{\!\!\restriction}

\renewcommand{\Re}{\mathop{\rm Re}\nolimits}
\renewcommand{\Im}{\mathop{\rm Im}\nolimits}

\theoremstyle{plain}

\newtheorem{thm}{Theorem}

\theoremstyle{definition}

\newtheorem{rem}{Remark}[section]

\numberwithin{equation}{section}

\def\bbbone{{\mathchoice {1\mskip-4mu {\rm{l}}} {1\mskip-4mu {\rm{l}}}
{ 1\mskip-4.5mu {\rm{l}}} { 1\mskip-5mu {\rm{l}}}}}

\def\squarebox#1{\hbox to #1{\hfill\vbox to #1{\vfill}}}

\newcommand{\tA}{\widetilde A}

\newcommand{\tkappa}{\tilde \kappa}

\newcommand{\tD}{\widetilde D}
\newcommand{\tE}{{\widetilde E}}

\newcommand{\tV}{\tilde V}

\newcommand{\eps}{\epsilon}

\def\t2{{\mathbb T}^2}

\newcommand{\cH}{{\mathcal H}}

\newcommand{\cK}{{\mathcal K}}
\newcommand{\cL}{{\mathcal L}}
\newcommand{\cM}{{\mathcal M}}

\newcommand{\cO}{{\mathcal O}}
\newcommand{\cP}{{\mathcal P}}
\newcommand{\cR}{{\mathcal R}}

\newcommand{\cS}{{\mathcal S}}
\newcommand{\cT}{{\mathcal T}}

\newcommand{\cU}{{\mathcal U}}

\newcommand{\im}{\operatorname{Im}}
\newcommand{\re}{\operatorname{\rm Re}}

\renewcommand{\Re}{\mathop{\rm Re}\nolimits}
\renewcommand{\Im}{\mathop{\rm Im}\nolimits}

\def\hto0{\xrightarrow{h\to 0}}

\usepackage{amsxtra}

\ifx\pdfoutput\undefined
  \DeclareGraphicsExtensions{.pstex, .eps}
\else
  \ifx\pdfoutput\relax
    \DeclareGraphicsExtensions{.pstex, .eps}
  \else
    \ifnum\pdfoutput>0
      \DeclareGraphicsExtensions{.pdf}
    \else
      \DeclareGraphicsExtensions{.pstex, .eps}
    \fi
  \fi
\fi

\title{Quantum transfer operators and quantum scattering}
\author%[S. Nonnenmacher]
{St\'ephane Nonnenmacher}
\address{Institut de Physique Th\'eorique\\
CEA/DSM/PhT (URA 2306 du CNRS)\\CE-Saclay\\91191 Gif-sur-Yvette\\ France}
\address{Institute of Advanced Study\\ Princeton, NJ 08540\\USA}
\email{snonnenmacher@cea.fr}
\begin{document}    

\maketitle   

%\tableofcontents

\section{Introduction and statement of the result}
\label{int}

\begin{figure}
\begin{center}
\includegraphics[width=0.35\textwidth]{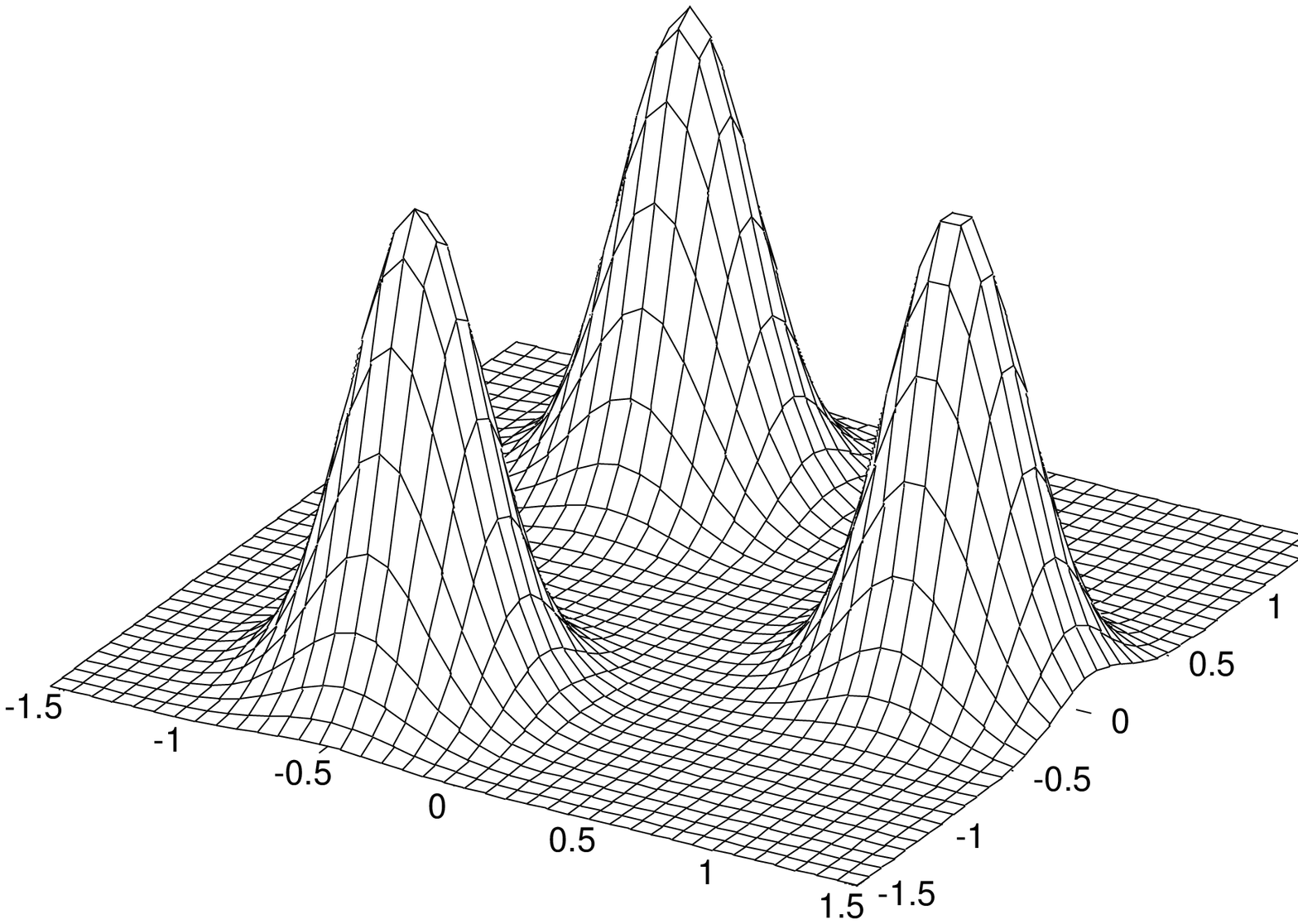}\hspace{1cm}
\includegraphics[width=0.55\textwidth]{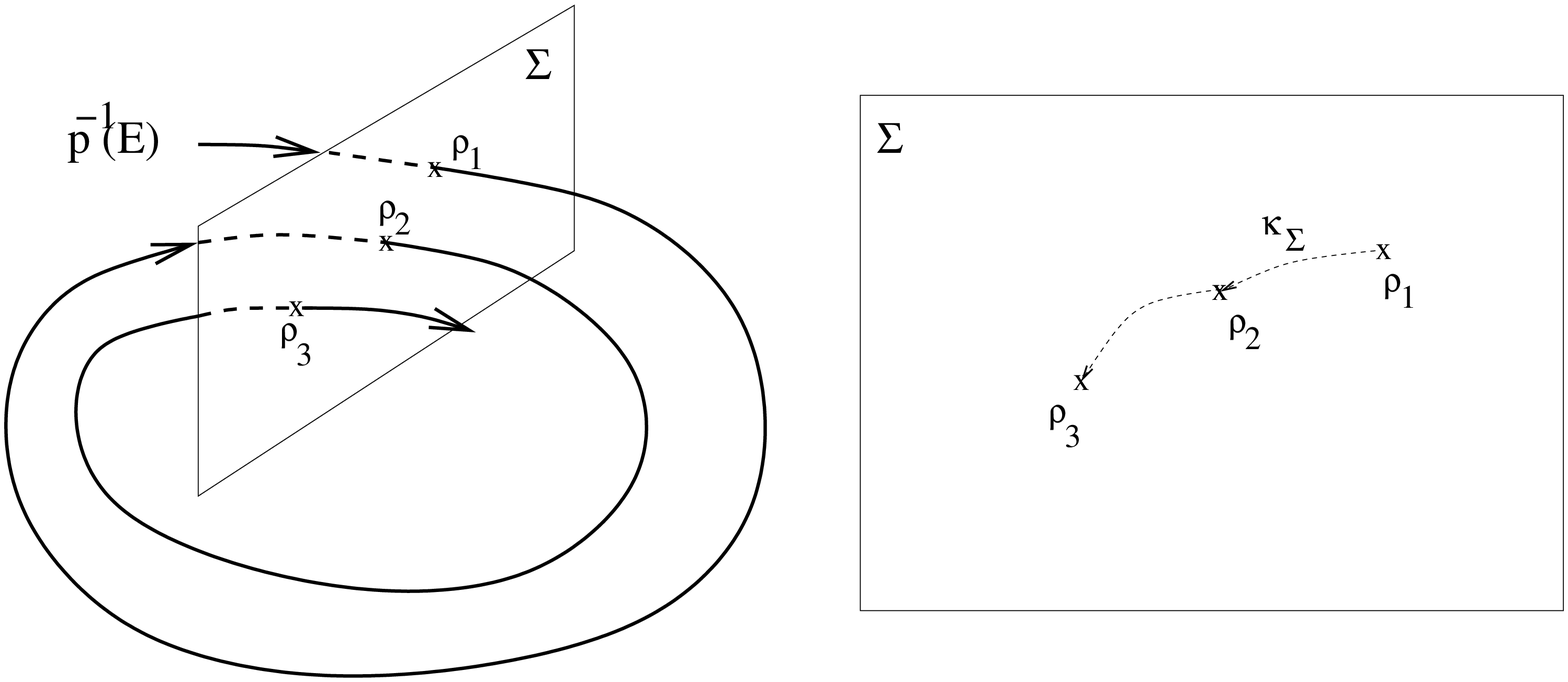}
\caption{Left: a 3-bump potential, which admits a fractal hyperbolic trapped set at intermediate energies \cite[Appendix]{SjDuke90}.
Right: schematic representation of a Poincar\'e section.\label{fig:pot}}
\end{center}
\end{figure}

These notes present a new method, developed in collaboration with 
Johannes Sj\"ostrand and Maciej Zworski, the aim of which is a better understanding of quantum
scattering systems, in situations where the set of classically
trapped trajectories at some energy $E>0$ is bounded, but can be a complicated fractal set.
In particular,
we are interested in the situations where this {\em trapped set} is a ``chaotic repeller'' hosting a hyperbolic
(Axiom A) flow. Such a scattering system belongs to the realm of {\em quantum chaos}, namely the study of
wave or quantum systems, the classical limit of which enjoy chaotic properties. This type of dynamics occurs for instance in
the scattering by $3$ or more disks in the Euclidean plane \cite{Ik}, but also in scattering by a smooth potential
(see fig \ref{fig:pot}). Chaotic scattering systems are physically relevant: for instance, mesoscopic
{\em quantum dots} are often modelled by open chaotic billiards \cite{NakaHara04};
the ionization of atoms or molecules in presence of external electric and/or magnetic fields
also involves classical chaotic trajectories \cite{BluRein97};
%. Still, the true classical dynamics in such systems is generally rather complicated, containing both ``regular parts'' 
%of phase space, embedded in a ``chaotic sea''. 
Open quantum billiards can also be realized in 
microwave billiard expermiments \cite{Stoe09}.

The method we propose is a quantum version of the Poincar\'e section/Poincar\'e map construction
used to analyze the classical flow (see \S\ref{s:Poincare0}). Namely, around some scattering energy $E>0$ 
we will construct a {\em quantum transfer operator}
(or quantum {\em monodromy} operator), which contains the relevant 
information of the quantum dynamics at this energy, in a much reduced form: this operator has finite rank
(which increases in the semiclassical limit), it allows to characterize the {\em quantum resonances}
of the scattering system in the vicinity of the energy $E$. The quantum transfer operator is very similar with the
{\it open quantum maps} studied as toy models for chaotic scattering \cite{casati,NZ1,schomerus}.

Our main result (Theorem \ref{thm:main}) will be stated in \S\ref{s:result}.
In \S\ref{s:formal} and \S\ref{s:well-posed} we sketch the proof of this result. We defer
the details of the proofs, as well as some applications of
the method, to a forthcoming publication \cite{NSZ1}.

\subsection*{From flows to maps, and back}\label{s:Poincare0}
Let us recall some facts from classical dynamics.
In the theory of dynamical systems, the study of a flow $\Phi^t:Y\to Y$ generated by some vector field (or ODE)
on a phase space $Y$ (say, a smooth manifold) can obten be facilitated
by considering a {\em Poincar\'e section} of that flow, namely a family 
$\Sigma=\{\Sigma_i,\,i=1,\ldots,J\}$ 
of hypersurfaces of $Y$, which intersect the flow transversely. The successive intersections
of the flow with $\Sigma$ define
a first return (or Poincar\'e) map $\kappa:\Sigma\to\Sigma$ (see fig. \ref{fig:pot}). 
This map, defined on a phase space $\Sigma$ of codimension $1$,
conveniently represents the flow on $Y$.
Long time properties of 
$\kappa$ are often easier to analyze than the corresponding properties of the 
flow. One can reconstruct the flow $\Phi^t$ from the knowledge of $\kappa$ together with the
{\em return time function} $\tau:\Sigma\to \RR_+$, which measures the time spanned between
the intersections $\rho$ and $\kappa(\rho)$. Below we will explain how {\em transfer operators} associated with
$\kappa$ can also help to compute long time properties of the flow.

%\begin{figure}
%\begin{center}
%\includegraphics[width=0.8\textwidth]{/Users/nonnen/images/poincare-section.eps}
%\end{center}
%\end{figure}
%The ergodic properties
%of the flow are related with those of $\kappa$, but also depend on the return function $\tau$.
%For instance, it has been known for almost fourty years that Anosov (that is, uniformly hyperbolic)
%smooth diffeomorphisms $\kappa$ are {\it exponentially mixing} with respect to the 
%natural invariant measure. On the other hand,
%an Anosov flow can be non-mixing, or
%mix very slowly. Specific conditions on the return map $\tau$ (like the 
%{\it uniform non-integrability condition} coined by Chernov) have been investigated more
%recently, which induce exponential mixing of the corresponding flow.

\subsubsection{Hamiltonian scattering}
The flows we consider are Hamiltonian flows defined on the cotangent space
$T^*\RR^n$. A Hamiltonian (function)
$p\in \CI(T^*\RR^n)$ defines a Hamilton vector field $H_p$ on phase space, which generates the flow $\Phi^t=\exp(tH_p)$.
For our specific choice \eqref{e:Hamilton}, the flow is complete.
%$$
%\binom{\dot{x}}{\dot{\xi}} =
%\binom{\frac{\partial p}{\partial \xi}}{-\frac{\partial p}{\partial x}}\defeq  H_p(x,\xi)\,,
%\quad(x,\xi)\in T^*\RR^n\,.
%$$
It preserves the symplectic structure on $T^*\RR^n$, and leaves invariant each energy shell $p^{-1}(E)$, so it makes
sense to study the dynamics on each individual shell.
A Poincar\'e section $\Sigma\subset p^{-1}(E)$ naturally inherits a 
symplectic structure, which is preserved by the Poincar\'e map $\kappa$. Hence, the Poincar\'e
maps we consider are (local) symplectomorphisms on $\Sigma$.

We will specifically consider Hamiltonians of the form 
\be\label{e:Hamilton}
p(x,\xi)=\frac{|\xi|^2}{2}+V(x),
\ee
with a potential $V\in \CIc(\RR^n)$ (say, supported in a ball $B(0,R_0)\subset\RR^n$). Such a Hamiltonian
generates a {\em scattering system}: for any energy $E>0$, particles can come from infinity, scatter on the 
the potential, and be sent back towards infinity. 
Depending on the shape of $V$ and of the energy, some trajectories can also be trapped forever (in the past
and/or in the future) inside the ball $B(0,R_0)$. This leads to the definition of the {\it trapped set} at energy $E$:
\be\label{e:K0} 
K_E \defeq \{\rho\in p^{-1}(E) \; : \;  \exp(\RR H_p ) (\rho) \text{ is 
bounded} \}, 
\ee
which is a compact, flow-invariant subset of $p^{-1}(E)$. The interesting long time dynamics takes place in the vicinity
of $K_E$, so the Poincar\'e section $\Sigma$ need only represent correctly the flow $\Phi^t$ restricted on
$K_E$, or on some neighbourhood of it. The Poincar\'e map $\kappa$ will also be defined in some neighbourhood of
the {\it reduced trapped set} $\cT_E\defeq K_E\cap\Sigma$. 
We will give a more precise description of $\Sigma$ in \S\ref{s:Poincare}.

\subsubsection{Chaotic dynamics and transfer operators}\label{s:transfer}

Our Theorem \ref{thm:main} will be relevant to the case where the flow on $K_E$ is uniformly hyperbolic (and satisfies
Smale's Axiom A). Such a flow is, in a sense ``maximally chaotic''. Hyperbolicity means that at each point $\rho\in K_E$
the tangent space $T_\rho p^{-1}(E)$ can be split between the flow direction, an unstable and a stable subspaces:
\be\label{e:AxiomA}
T_\rho p^{-1}(E)=\RR H_p\oplus E^+(\rho)\oplus E^-(\rho)\,,
\ee
where the (un)stable subspaces are defined by the long time properties of the tangent map: there exist $C,\lambda>0$ such that,
for any $\rho\in K_E$,
$$
v\in E^\mp(\rho)\Longleftrightarrow \|d\Phi^{\pm t} v\|\leq C\,e^{-\lambda t},\quad t>0\,.
$$
The Poincar\'e map $\kappa$ then inherits the Axiom A property.
%the reduced trapped set $\cT_E$.

To study the long time properties of such chaotic flow, it has proved convenient to use {\it transfer operators}
associated with $\kappa$ \cite{Baladi}. Let us give an example of such operators.
Given any weight function $f\in C(\Sigma,\RR)$, 
one define the transfer operator $\cL_f$ by a weighted push-forward on functions $\varphi:\Sigma\to \RR$:
$$
\cL_f\, \varphi (\rho) \defeq \sum_{\rho':\kappa(\rho')=\rho}e^{f(\rho')}\,\varphi(\rho')\,.
$$
Provided $\cL_f$ is applied to some appropriate functional space\footnote{The functional space can be rather complicated, see e.g. \cite{GouLiv06} for the case of Anosov diffeomorphisms.}, its {\it spectrum} can deliver relevant information about the long time dynamics of $\kappa$.
For instance, the spectral radius $r_{sp}(\cL_f)$
determines the {\it topological pressure} of $\kappa$ associated with the weight $f$, 
which provides statistical information on the long periodic orbits of $\kappa$:
$$
\log r_{sp}(f)=\cP(\kappa,f)\defeq \lim_{T\to\infty}\frac1T\log\sum_{|\gamma|\leq T} e^{\int_\gamma f}\,.
$$
(here $\int_\gamma f$ is the sum of values of $f(\rho)$ along the periodic orbit $\gamma$).
The topological pressure of the {\em flow} $\Phi^t$, associated with
a weight $F\in C(X)$, can also be computed through transfer operators. One defines on $\Sigma$ the function 
$f(\rho)=\int_0^{\tau(\rho)}F(\Phi^t(\rho))$, that is
the accumulated weight
from $\rho\in\Sigma$ to its next return $\kappa(\rho)$, and considers the family $\{\cL_{f-s\tau},\ s\in\RR\}$ of transfer
operators. The following relation then relates the pressures
of $\kappa$ and $\Phi^t$:
$$
s=\cP(\Phi^t,F)\Longleftrightarrow \cP(\kappa,f-s\tau)=0\Longleftrightarrow r_{sp}(\cL_{f-s\tau})=1\,.
$$
The decay of correlations for the Axiom A flow $\Phi^t$ is encoded in the {\it Ruelle-Pollicott resonances}, which are the 
poles of the Fourier transform of the correlation function \cite{Pollicott85}. Within some strip $\cS\subset\CC$, 
these resonances $\{z_i\}$ can be characterized by using the family $\{\cL_{f-z\tau},\ z\in\CC\}$ 
of complex weighted transfer operators: 
$z_i\in\cS$ is a resonance iff $\cL_{f-z_i\tau}$ has an eigenvalue equal
to $1$. This property can be written (abusively, because transfer
operators are usually not trace class) as
\be\label{e:det-classical}
z\in\cS\ \text{is a Ruelle-Pollicott resonance}\Longleftrightarrow \det(1-\cL_{f-z\tau})=0\,.
\ee

\subsection{A quantum scattering problem}
Let us now introduce the quantum dynamics we are interested in.
The operator 
$$ 
P = P(h)= - \frac{h^2 \Delta}{2} + V( x )  \,, \quad V \in \CIc ( \RR^n ) \,,
$$
generates the Schr\"odinger dynamics $U(t)=\exp(-itP(h)/h)$ on $L^2(\RR^n)$. $P(h)$
is the $h$-quantization of the classical Hamiltonian \eqref{e:Hamilton}, so the semiclassical behaviour of the 
quantum dynamics will be strongly influenced by the flow $\exp(tH_p)$.
We focus on the dynamics around some positive energy $E>0$, so the flow we need to understand is $\Phi^t\rest p^{-1}(E)$. 
We will assume that
\begin{itemize}
\item the flow on $p^{-1}(E)$ has no fixed point:  $ dp \rest_{p^{-1} ( E ) } \neq 0 $.
\item the trapped set $K_E$ has topological dimension $1$. Equivalently, the reduced trapped set
$\cT_E=K_E\cap \Sigma$ is totally disconnected.
\end{itemize}
These conditions are satisfied, for example, for a 3-bump potential at intermediate energies (see fig. \ref{fig:pot}).
The second condition was absent in previous studies of such systems \cite{SjZw04,NZ2}, it is a technical
constraint specific to the approach we develop below (as we explain after Thm \ref{thm:main}, the condition
required for the method to work is actually weaker).

We are interested in the long time Schr\"odinger dynamics near
energy $E$, so it is natural to investigate the spectrum of $P(h)$ near $E$. That operator is self-adjoint on $L^2(\RR^n)$, 
with domain $H^2(\RR^n)$; but, due to the bounded support of $V(x)$, the spectrum of $P(h)$ is absolutely continuous on 
$\RR^+$, without any embedded eigenvalue. 
%This continuous spectrum reflects the fact that energy shells $p^{-1}(E)$ are unbounded for any $E>0$.
Nevertheless, the truncated resolvent 
$\psi (P(h)-z)^{-1}\psi$ (with $\psi\in\CIc(\RR^n)$), well-defined in the quadrant $\{\re z>,\,\im z>0\}$,
can be meromorphically extended across the real axis into $\{\re z>0,\,\im z<0\}$. The finite rank poles 
$\{z_i\}$ in this region are called its {\em resonances} (they do not depend on the specific cutoff $\psi$). 
Resonances are often understood as ``generalized eigenvalues'':
they are associated with {\it metastable} modes $u_i(x)$ which are not square-integrable,
but satisfy the differential equation $P(h)u_i = z_i\,u_i$, so that they decay expontially in time, at a rate given by $|\im z_i|/h$.

\begin{figure}
\begin{center}
\includegraphics[width=.6\textwidth]{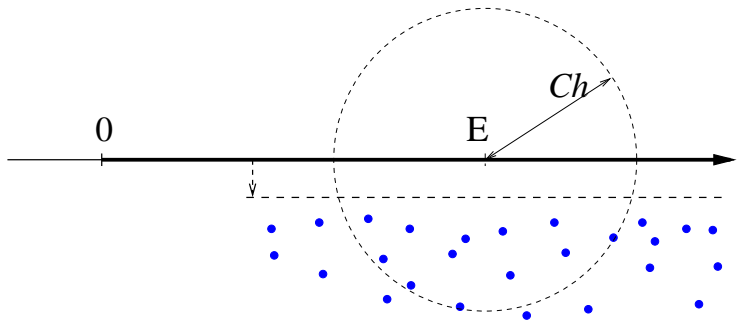}
\caption{\label{fig:resos1p} Schematic representation of the spectrum of $P(h)$ and its resonances near the energy $E$.}
\end{center}
\end{figure}

One of our objectives is to better understand the distribution of these
resonances in the 
$h$-neighbourhood of the energy $E$, that is in disks $D(E,Ch)$ (see fig. \ref{fig:resos1p}). More precisely, 
we want to investigate:
\begin{itemize}
\item the number of resonances in $D(E,Ch)$. So far {\em fractal upper bounds} have been proven \cite{SjZw04}. We wish to
investigate whether similar lower bounds can be obtained, at least for a generic system. 
\item the width of the resonance free strip in $D(E,Ch)$. A lower bound for such a strip has been
expressed in terms of some topological pressure \cite{Ik,GasRice,NZ2}, but a recent result of Petkov-Stoyanov
(for obstacle scattering) shows that this lower bound is in general not sharp \cite{PetStoy07}.
\end{itemize}

\subsection{Our result}\label{s:result}
Our main result is a  ``quantization'' of the Poincar\'e section method
presented above.
\begin{thm}\label{thm:main}
Assume that, for some energy $E>0$, the trapped set $K_E$ for the flow $\exp(tH_p)$ is topologically one dimensional,
and contains no fixed point.

Then, for $h>0$ small enough, there exists a family of matrices $ \{M ( z , h ),\,  z \in D ( 0 , Ch )\}$
holomorphic w.r.to $z$,
such that the zeros of the function
\be\label{e:det-quantum}
\zeta ( z , h ) \defeq \det ( I - M ( z , h ) ) 
\ee
give the resonances of $ (P(h)-E) $ in $ D (0, Ch ) $, with correct multiplicities.

The matrices $M ( z , h )$ have the following structure. 
There exists a Poincar\'e section $\Sigma=\sqcup_{i=1}^J\Sigma_i$ and map $\kappa:\Sigma\to\Sigma$, an $h$-Fourier integral operator 
$\cM ( z , h ):L^2(\RR^{n-1})^J\circlearrowleft$ quantizing $\kappa$,
and a projector $\Pi_h$ of rank $r(h)\asymp h^{-n+1}$, such that
$$
M ( z , h ) =  \Pi_h\, \cM ( z , h )\, \Pi_h  + \cO(h^N ) \,.
$$
The remainder estimate holds in the operator norm on $\CC^{r(h)}$. The exponent $N$ can be assumed arbitrary large.
\end{thm}

\begin{rem}
The $1$-dimensional condition we impose on $K_E$ is not strictly necessary. What one needs
is the existence of a Poincar\'e section $\Sigma$ intersecting $\Phi^t\rest K_E$, 
such that $\partial\Sigma\cap K_E=\emptyset$; in particular, we don't need the flow to be hyperbolic on 
$K_E$. Still, Axiom A flows provide the most obvious example for which this condition is satisfied \cite{bowen-1dim-flows}; 
it holds as well for the broken geodesic flow in the scattering by $3$ disks satisfying a no-eclipse condition \cite{Ik}.
\end{rem}
This theorem shows that the dynamics generated by the Hamiltonian
$P(h)$ near $E$ can be ``summarized'' in the 
family of {\it quantum transfer operators} $\{M(z,h),\,z\in D(0,Ch)\}$. One reason for this terminology
is that $M(z,h)$ bears some resemblance with the
transfer operators $\cL_{f-z\tau}$ briefly described in \S\ref{s:transfer}. 
The equation \eqref{e:det-quantum}
characterizing quantum resonances is obviously the quantum analogue of the (generally formal) equation \eqref{e:det-classical}
defining Ruelle-Pollicott resonances. Also, the notation $\zeta(z,h)$ in \eqref{e:det-quantum}
hints at an analogy, or relationship, between this spectral determinant and some form of {\it semiclassical zeta function}
(such functions have been mostly studied in the physics literature, see e.g. \cite{CvRosVatRug93}).

The operators $M(z,h)$ have the same semiclassical structure as {\it open quantum maps} studied
in the (mostly physical) literature as toy models of quantum scattering systems. For instance, the distribution
of resonances and resonant modes 
has proven to be much easier to study numerically for open quantum maps, than for realistic flows 
\cite{casati,schomerus,NZ1,KNPS06}.
The novelty here, is that the operators $M(z,h)$ allow to characterize a ``physical'' resonance spectrum.

\subsubsection{Historical remarks}
Actually, a similar method has been introduced in the theoretical physics
literature devoted to ``quantum chaos''. 
To the author's knowledge, the first such construction appeared in Bogomolny's work
\cite{Bogo} on multidimensional closed quantum systems. In that work, a family of 
quantum transfer operators $T(E)$ is
constructed, which are integral operators defined on a hypersurface in configuration space. The eigenvalues of the
bound Hamiltonian are then obtained, in the semiclassical limit, as roots of the equation $\det(1-T(E))=0$. 
This work generated a lot of interest in the quantum chaos community.
Smilansky and co-workers derived a similar quantization condition for closed Euclidean 2-dimensional billiards 
\cite{DorSmil92}, replacing $T(E)$ by a scattering matrix $S(E)$
associated with the dual scattering problem. Bogomolny's
method was also extended to study quantum scattering situations \cite{GeorPran95}.
On the other hand, Prosen developed an ``exact'' (that is, not necessarily semiclassical) 
quantum surface of section method to study certain closed Hamiltonian systems \cite{Prosen94}.

In the mathematics literature
similar operators appeared in the framework of obstacle scattering \cite{Ge,Ik}: the 
scattering problem was analyzed through integral operators defined on
the obstacle boundaries, which also have the structure of Fourier integral operators associated with
the bounce map.
More recently, a {\it monodromy operator} formalism has been introduced in \cite{SjZw02} to study the Schr\"odinger
dynamics in the
vicinity of a single isolated periodic orbit. This approach has then been used to investigate
concentration properties of eigenmodes in the vicinity of such an orbit \cite{Chris09}. 
The construction we present below heavily borrows from the techniques developed in
\cite{SjZw02}. It improves them on two aspects: first, our invariant
set $K_E$ is more complex than a single periodic orbit. Second, the connection we establish between the operators
$(P(h)-E-z)$ and $M(z,h)$ is deeper than previously.

\section{Formal construction of the quantum transfer operator}\label{s:formal}

The proof of Thm \ref{thm:main} proceeds in several steps. It uses many tools of $h$-pseudodifferential 
calculus (we will use the notations of \cite{DiSj,EZ}). We just recall a few of them:
\begin{itemize}
\item a state $u=u(h)\in L^2$ is microlocalized in a domain $U\Subset T^*\RR^n$ iff, 
for any function
$\chi\in \CIc(T^*\RR^n)$ with $\supp \chi\cap \bar U=\emptyset$, one has $\|\chi^wu\|_{L^2}=\cO(h^\infty)\|u\|_{L^2}$. 
(here $\chi^w=\chi^w(x,hD_x)$ denotes the $h$-Weyl quantization of $\chi$).
\item two states $u,\,v$ are said microlocally equivalent in $U\Subset T^*\RR^n$ iff, for any cutoff $\chi\in\CIc(U)$,
one has $\|\chi^w(u-v)\|_{L^2}=\cO(h^\infty)$.
\item similarly, two operators $A,B$ are said microlocally equivalent
in $V\times U$ (with $U,V\Subset T^*\RR^n$) iff, for any cutoffs $\chi_1\in\CIc(U)$, $\chi_2\in\CIc(V)$, one has 
$\|\chi_2^w(A-B)\chi^w_1\|_{L^2\to L^2}=\cO(h^\infty)$. 
\item an operator $A$ is microlocally defined in $V\times U$ iff it is microlocally equivalent in $V\times U$
to some globally defined operator. ``Microlocally defined in $U$'' will mean ``microlocally defined in $U\times U$''.
\end{itemize}
The present section constructs the quantum transfer operators microlocally in a neighbourhood of the trapped set $K_E$, 
without paying attention to the rest of the phase space. 
The arguments making the construction globally well-defined will be presented in \S\ref{s:well-posed}. 

The microlocal construction being strongly tied to a Poincar\'e section, we start by describing the latter in some detail.

\subsection{Description of the Poincar\'e section}\label{s:Poincare}
%\begin{figure}
%\begin{center}
%\includegraphics[width=.6\textwidth]{/Users/nonnen/zworski/Grushin/cutoffs2bis.eps}
%\caption{Schematic representation of the neighbourhood of the trapped set $K_E$ (green), with a few
%trapped trajectories. Poincar\'e 
%sections are represented by dashed lines, and the ``arrival'' subsets $\tA_{ij}$ by thick red lines.
%We also show the supports of the cutoffs $\chi_i^b$ (``pants'') and $\chi_i^f$ (rectangles) 
%used in the definition of our Grushin problem (see section \ref{s:Grushin}).\label{f:cutoffs2bis}}
%\end{center}
%\end{figure}

According to the assumptions of the theorem, the trapped set $K_E$ is a compact set of topological dimension unity. 
It is then possible to construct a Poincar\'e section 
$\Sigma=\sqcup_{i=1}^J\Sigma_i\subset p^{-1}(E)$ with the following properties:
\begin{itemize}
\item each $\Sigma_i$ is a $(2n-2)$-dimensional topological disk, transverse to the flow. 
\item the maximal diameter of the $\Sigma_i$ can be chosen arbitrary small.
\item there exists a time $\tau_{\max}>0$ such that, for any $\rho\in K_E$, the trajectory $\Phi^t(\rho)$ intersects
$\Sigma$ at some time $0<t\leq \tau_{\max}$. 
\item the boundary $\partial \Sigma=\sqcup_i\Sigma_i$ does not intersect $K_E$. 
\end{itemize}
If we restrict ourselves to points in the reduced trapped set $\cT_E\defeq K_E\cap \Sigma$, 
the map $\rho\mapsto \rho_+(\rho)$ defines a bicontinuous bijection $\kappa:\cT\to\cT$.

\begin{figure}
\begin{center}
\includegraphics[width=.7\textwidth]{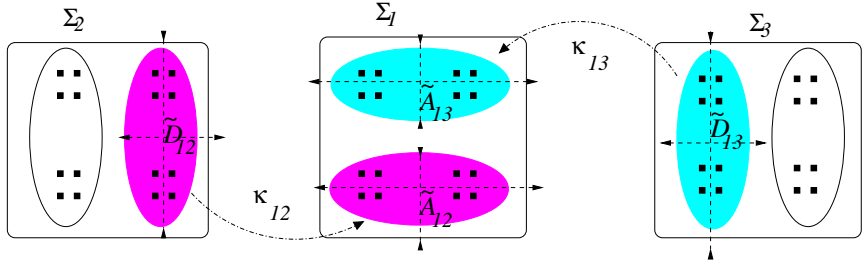}
\caption{\label{fig:sections3-}Schematic representation of a hyperbolic Poincar\'e map.
The light blue (resp. pink) regions represent $\tD_{31}$ and $\tA_{31}$ (resp.  $\tD_{21}$ and $\tA_{21}$).
The (un)stable directions are represented by the dashed horizontal and vertical lines. The black squares 
show a coarse-graining of $\cT_E$.}
\end{center}
\end{figure}

Each component of the reduced trapped set, $\cT_{i}\defeq K_E\cap\Sigma_i$, splits in two different ways: 
\begin{align}
\cT_i=\sqcup_j D_{ji},\quad &\text{where}\quad D_{ji}=\{\rho\in \cT_i,\ \kappa(\rho)\in \cT_j\}\\
\cT_i=\sqcup_j A_{ij},\quad &\text{where}\quad A_{ij}=\{\rho\in \cT_i,\ \kappa^{-1}(\rho)\in \cT_j\}
\end{align}
We will denote by $J_+(i)$ (resp. $J_-(i)$) the set of indices in the ``outflow'' (resp. ``inflow'') of $\cT_i$, that is such that 
$D_{ji}$ and $A_{ji}$ (resp. $D_{ij}$ and $A_{ij}$) are not empty.
The map $\kappa$ is the union of components $\kappa_{ij}$, which relate bijectively $D_{ij}$ with $A_{ij}$.
Since $\cT_i$ lies in the interior of $\Sigma_i$, the components $D_{ji}$ (resp. $A_{ij}$) are disconnected
from one another. 
Hence, each $\kappa_{ij}$ can be extended to be a bijection $\kappa_{ij}:\tD_{ij}\to \tA_{ij}$, 
where $\tD_{ij}$, $\tA_{ij}$ are open neighborhoods of $D_{ij}$ and $A_{ij}$, respectively in $\Sigma_j$ and $\Sigma_i$.
The extended map $\kappa_{ij}:\tD_{ij}\to\tA_{ij}$ is a symplectomorphism (see fig. \ref{fig:sections3-} for a sketch).

\subsection{Microlocal solutions}\label{s:solutions}
In this section we show that, for $z\in D(0,Ch)$, any solution to the equation $(P(h)-E-z) u =0$, microlocally
near some part of $K_E$,
can be ``encoded'' by a transversal function $w\in L^2(\RR^{n-1})$, which ``lives'' on one component $\Sigma_i$
of the Poincar\'e section.

Take such a component $\Sigma_i$. From the assumption $dp\rest_{p^{-1}(E)}\neq 0$, 
there exists an open neighbourhood 
$V_i$ of $\Sigma_i$, and a set of symplectic coordinates
$(y_1,\ldots,y_n;,\eta_1,\ldots,\eta_n)$ on $V_i$, such that
\begin{itemize}
\item the Hamiltonian $p(\rho)=E+\eta_1$ for any $\rho\in V_i$
\item the section $\Sigma_i$ is locally defined by $\{y_1=\eta_1=0\}$, and the origin $y=\eta=0$ corresponds to
a point in $\cT_i$.
\end{itemize}
Equivalently, there exists a neighbourhood $(0,0)\in\tV_i\in T^*\RR^{n}$ and a 
symplectomorphism $\tkappa_i:\tV_i\to V_i$, such that $p\circ\tkappa_i(y,\eta)=E+\eta_1$, etc.

The change of coordinates $\tkappa_i$ can be $h$-quantized into
an $h$-Fourier integral operator $\cU_i:L^2(\RR^n)\to L^2(\RR^n)$, microlocally defined and unitary near 
$V_i\times\tV_i$, such that 
\be\label{e:normalform}
\cU_i^*\,(P(h)-E)\,\cU_i\quad\text{is microlocally equivalent with $hD_{y_1}$ in $\tV_i\times\tV_i$}.
\ee
This ``quantum change of coordinates'' allows one to easily characterize, for $z\in D(0,Ch)$,
the microlocal solutions to the equation
\be\label{e:(P-z)u}
(P(h)-E-z) u =0\qquad \text{microlocally in $V_i$}\,.
\ee
Indeed, the equation
\be\label{e:hD}
(hD_{y_1}-z) v=0
\ee
is obviously solved by
\be\label{e:wtov}
v(y_1,y')=e^{izy_1/h}\,w(y'),\quad w\in L^2(\RR^{n-1})\,,
\ee
that is by extending some ``transversal data'' $w$. We denote this extension by $v=\cK(z)w$. 
Conversely, the solution $v$ can easily be ``projected'' onto the data $w$: consider some monotone $\chi\in \CI(\RR^n)$ such 
that $\chi(y)=0$ for $y_1<\eps$, $\chi(y)=1$ for $y_1>\eps$. Then, $w$ can be recovered from $v$ through
$$
w(y')=\int_{\RR} e^{-izy_1/h}\,\partial_{y_1}\chi(y)\,v(y)\,dy_1\,.
$$
In a more compact form, we write $w=\cK(\bar z)^*\chi'\,v$, with $\chi'=\frac{i}{h}[hD_{y_1},\chi]$. 

The solutions of \eqref{e:(P-z)u} are then given by selecting some $w\in L^2(\RR^{n-1})$ (microlocalized near the origin),
and take
\be\label{e:Poisson}
u=\cU_i\cK(z)w\defeq \cK_{i}(z)\,w\,.
\ee
That is, the operator $\cK_i(z)$ builds a microlocal solution of \eqref{e:(P-z)u} near
$\Sigma_i$, starting from ``transversal data'' $w\in L^2(\RR^{n-1})$. The latter can be interpreted as
a quantum state living in the reduced phase space $\Sigma_i$. The converse ``projection'' is given by 
\be\label{e:projection-i}
w=\cK(\bar z)^*\chi'\,\cU_i^*\,u=\cK_i(\bar z)^*\chi'_i\,u\defeq R_{+i}(z)\,u\,.
\ee
Here $\chi_i$ is the cutoff corresponding to $\chi$ near the section $\Sigma_i$. To get a consistent definition for $\chi_i$,
we must assume that it jumps back down to $0$ a little further along the flow, but the precise position will be
irrelevant. Indeed, the commutator
$\frac{i}{h}[P(h),\chi^w_i(x,hD_x)]$ is equal (microlocally near $K_E$) to the sum of two pseudodifferential operators with
disjoint wavefront sets.
The first one is microlocalized near $\Sigma_i$ (in the region where
$\chi_i$ jumps from $0$ to $1$), we will denote it by
$\chi_i'=\frac{i}{h}[P(h),\chi^w_i(x,hD_x)]_i$; the second component ``lives'' in the region where 
$\chi_i$ decreases from $1$ to $0$, and will not play any role.

The same construction can be performed independently near each $\Sigma_j$, $j=1,\ldots,J$. We will call $w_j$ the 
transversal data associated with the section $\Sigma_j$, and $\cK_j(z)$, $R_{+j}(z)$ the corresponding operators.

\subsection{From one transversal parametrization the next}
The the solution \eqref{e:Poisson} is microlocalized in $V_i$, since $\cU_i$ is only defined microlocally
in $V_i\times\tV_i$. However, this solution can be extended in a forward cylinder $\cup_{0\leq t\leq T}\Phi^t\Sigma_i$ by using the 
propagator $e^{-it(P-E-z)/h}$: if $u$ is a solution near $\rho\in\Sigma_i$, then $e^{-it(P-E-z)/h}u$ is
the extension of this solution near $\Phi^t(\rho)$.

This way, we can extend $u$ up to the vicinity of the sections $\Sigma_j$ in the outflow of $\Sigma_i$.
This extended solution will still be denoted by $u=\cK_i(z)w_i$. Near $\Sigma_j$, this solution can also be parametrized
by the ``transversal'' function $w_j=R_{+j}\, u\in L^2(\RR^{n-1})$.
The map $w_i \mapsto w_j$, which amounts to changing the transversal parametrization for a single solution $u$,
defines our quantum Poincar\'e map:
\be
\cM_{ji}(z,h)\defeq R_{+j}(z)\cK_i(z)\,.
\ee 
This operator is a Fourier integral
operator quantizing the Poincar\'e map $\kappa_{ji}$; it is 
microlocally defined, and microlocally unitary, on $\tD_{ji}\times \tA_{ji}$.

Let us see how the operators $\cM_{ji}(z)$ can be used. Assume $E+z$ is a resonance of $P(h)$, with $z\in D(0,Ch)$. Then,
there exists a metastable state $u\in L^2_{loc}$, global
solution to the equation $(P-E-z)u=0$. The above procedure associates to this solution $J$ parametrizations
$w_i=R_{+i}(z) u$, microlocally defined near $\cT_j$. For any $i$ and $j\in J_+(i)$, these
parametrizations satisfy $w_j=\cM_{ji}(z)w_i$: can be written in the compact form
\be\label{e:identity}
w = \cM(z)\, w\,,
\ee
where $w = (w_i)_{i}$ is the column vector of all $J$ local parametrizations, and $\cM(z)$ is the
operator valued matrix $\big(\cM_{ij}(z)\big)$. 

In the next subsections we prove that the converse statement holds as well: the existence of a solution of $(\cM(z)-Id) w=0$ microlocally
near $\cT_E$ implies the existence of a solution of $(P-E-z)u=0$ microlocally near $K_E$.
To prove this we will set up a formal Grushin problem, in which the operator $(\cM(z)-I)$ will appear as the 
``effective Hamiltonian'' for the original operator $(P-E-z)$.

\subsection{Grushin problems}\label{e:Grushin-general}

A {\em Grushin problem} for the family of operators\footnote{$H^2_h(\RR^n)$ is the semiclassical Sobolev space of norm
$\|u\|_{H^2_h}=\int |\tilde u(\xi)|^2 (1+|h\xi|^2)^2\,d\xi$, with $\tilde u$ the Fourier transform of $u$.} 
$\{(P-E-z):H^2_h(\RR^n)\to L^2(\RR^n),\ z\in D(0,Ch)\}$ 
consists in the insertion of that operator inside an operator valued matrix
\be\label{e:Grushin0}
\cP(z)=\begin{pmatrix}\frac{i}{h}(P-E-z)&R_-(z)\\R_+(z)&0\end{pmatrix}: H^2_h(\RR^)\times\cH\to L^2(\RR^)\times\cH \,,
\ee
in a way such that $\cP(z)$ is invertible (see e.g. \cite{SZ9} or \cite[Appendix]{EZ} for a general presentation of this method). 
Ideally, the auxiliary space $\cH$ is ``much smaller'' than $L^2$ or $H^2_h$ (in our final version, $\cH$ will be finite dimensional). 
The inverse of $\cP(z)$ is traditionally written in the form
$$
\cP(z)^{-1}= \begin{pmatrix} E ( z ) & E_+ ( z ) \\
E_- (z )  &  E_{-+} ( z )  \end{pmatrix}\,.
$$
The invertibility of $(P-E-z)$ is then equivalent with that of the operator $E_{-+}(z)$: Schur's complement formula shows that
\be\label{e:inversion}
\frac{h}{i}(P-E-z)^{-1}=E(z)-E_+(z)E_{-+}(z)^{-1}E_-(z)\,,\qquad E_{-+}(z)^{-1}=-\frac{h}{i} R_+(z)(P-E-z)^{-1}R_-(z)\,,
\ee
so that $\dim\ker(P-E-z)=\dim\ker E_{-+}(z)$.
For this reason, $E_{-+}(z)$ is called an {\it effective Hamiltonian} associated with $(P(h)-E-z)$. It has a smaller
rank than $P(h)$, but its dependence in the spectral parameter $z$ is nonlinear.
%The Grushin problem we will setup below will have for effective Hamiltonian $E_{-+}(z)=\cM(z)-I$; hence,
%a nontrivial solution of $w=\cM(z)\,w$ is equivalent with the existence of a microlocal solution of $(P-E-z)u=0$ near $K_E$.

\subsection{Our formal Grushin problem}\label{s:Grushin}
We will first build our Grushin problem microlocally near $K_E$ (so we can identify $H^2_h$ with $L^2$).
Our auxiliary space $\cH$ will contain local ``transversal data'' $w_i\in L^2(\RR^{n-1})$, one for each section $\Sigma_i$,
so we have formally $\cH=L^2(\RR^{n-1})^J$.
The auxiliary operators are then vectors of operators:
$R_{+}(z)=(R_{+1},\ldots,R_{+J})$, $R_{-}(z)=\ ^t\!(R_{-1},\ldots,R_{-J})$, which will for now be defined only
microlocally:
\begin{itemize}
\item $R_{+i}(z)$ is the ``projector'' \eqref{e:projection-i} of $L^2(\RR^n)$ onto the parametrization $w_i$ living on $\Sigma_i$. 
We will rebaptize $\chi_i\defeq\chi_i^f$ (for {\it f}orward) the cutoff used in the definition of $R_{+i}$. 
\item on the opposite, $R_{-i}(z)$ takes the data $w_i\in L^2(\RR^{n-1})$ to produce a microlocal solution, and cuts off this solution by
applying the derivative of another cutoff $\chi_i^{b}$:
\be\label{e:R_-}
R_{-i}(z)=\chi_i^{b\prime}\cK_i(z)\,.
\ee
The cutoff $\chi_i^{b}$  (for {\it b}ackward) is  similar with $\chi_i^{f}$, and
$\chi_i^{b\prime}$ is, as before, the component of $[\frac{i}{h}P(h),(\chi_i^b)^w]$ microlocalized near $\Sigma_i$. 
We require that the jump of $\chi_i^b$ occurs {\em before} that of $\chi_i^f$, and that
the whole family $\{\chi_i^b,\ i=1,\ldots,J\}$ satisfies a local resolution of identity near $K_E$:
\be\label{e:resolution}
\sum_{i}\chi_i^b=1\,,\quad\text{in some neighbourhood of $K_E$}.
\ee
\end{itemize}
%The two types of cutoffs are sketched in the figure \ref{f:cutoffs2bis}.

\subsubsection{Homogeneous problem}\label{s:homog}
Let us now try to invert the matrix $\cP(z)$ we have just defined, at least microlocally near 
$K_E\times \prod_i\cT_i$. 
First we consider arbitrary transversal data $ w=(w_i)$, and try to solve (in $u\in L^2(\RR^n)$,
$ u_{-}\in L^2(\RR^{n-1})^J$) the system
\begin{align}
\label{e:1}\frac{i}{h}(P-E-z)u+\sum_{i=1}^J R_{-i}(z)u_{-i}&=0\\
\label{e:2}R_{+i}(z)u&=w_{i},\quad i=1,\ldots,J\,.
\end{align}
Eq.~\eqref{e:2} suggests that $u$ could be a microlocal solution parametrized by $w_{i}$, at least
in the region where $\chi_j^{f}$ jumps from $0$ to $1$. Since $\chi^b_i\equiv 1$ 
in this region, we take the Ansatz
\be\label{e:u-homog}
u=\sum_{i=1}^J  (\chi_i^b)^w\cK_i(z)\,w_i\defeq \sum_i E_{+i}(z)\,w_i\,.
\ee
Injecting this Ansatz in \eqref{e:1}, we obtain
\be\label{e:homog2}
\sum_{i=1}^J \frac{i}{h}[P, (\chi_i^b)^w]\cK_i(z)\,w_i
+\sum_{i=1}^J R_{-i}(z)u_{-i}=0\,,
\ee
which we want to solve in $(u_{-i})$.
Each commutator $\frac{i}{h}[P, (\chi_i^b)^w]$ is the sum of 
a component $\chi_i^{b\prime}=\frac{i}{h}[P, (\chi_i^b)^w]_i$ microlocalized
near $\Sigma_i$, and of components $\frac{i}{h}[P, (\chi_i^b)^w]_j$ microlocalized near 
$\tA_{ji}\subset \Sigma_j$, for each index $j\in J_+(i)$. 
The resolution of identity \eqref{e:resolution} shows that near  we have
$H_p\chi_i^{b\prime}+H_p\chi_j^{b\prime}=0$,  the quantum version of which reads 
$\frac{i}{h}[P, (\chi_i^b)^w]_j+\frac{i}{h}[P, (\chi_j^b)^w]_j=0$ microlocally near $\tA_{ji}$. As a result \eqref{e:homog2} can be
rewritten as
$$
\sum_{i=1}^J \chi_i^{b\prime}\cK_i(z)\,w_i
-\sum_{i=1}^J \sum_{j\in J_+(i)} \chi_j^{b\prime}\cK_i(z)\,w_i
+\sum_{i=1}^J R_{-i}(z)u_{-i}=0\,.
$$
Near each $\Sigma_j$, $j\in J_+(i)$ we have $\cK_i(z)w_i=\cK_j(z)\cM_{ji}(z)\,w_i$. For each $i$ we can group 
together the terms localized near $\Sigma_i$, and get:
$$
R_{-i}(z)\,w_i -\sum_{i\in J_+(j)} R_{-i}(z)\cM_{ij}(z)\,w_j
+R_{-i}(z)u_{-i}=0\,.
$$
This leads to the microlocal solution
\be\label{e:u_{-i}}
u_{-i}=-w_i + \sum_{i\in J_+(j)}\cM_{ij}(z)\,w_j\defeq \sum_{j}E_{-+ij}(z)w_j\,.
\ee
We have thus solved the system (\ref{e:1},\ref{e:2}) microlocally near $K_E\times \prod_i\cT_i$, and provided
explicit expressions for the operators $E_+(z)$ and $E_{-+}(z)=\cM(z)-Id$, microlocally near the trapped set.

\subsubsection{Nonhomogeneous problem}
To complete the microlocal inversion of $\cP(z)$, we now take $v\in L^2(\RR^n)$ microlocalized near $K_E$, 
and try to solve (in $u,\,u_{-}$, microlocally near $K_E$)
\be
\label{e:1'}\frac{i}{h}(P-E-z)u+\sum_{i=1}^J R_{-i}(z)u_{-i}=v\,.
\ee
Let us first assume that $v$ is microlocalized inside the region $\{\chi_i^b(\rho)=1\}$ for some index $i$.
We then take the truncated parametrix 
$\tE(z)=\int_0^T e^{-it(P-E-z)/h}\,dt$, with $T$ large enough so that $e^{-iTP/h}v$ is microlocalized 
beyond $\supp\chi^b_i$, and define the Ansatz $u=(\chi^b_i)^w\,\tE(z)\,v$.
The latter satisfies
\begin{align}\frac{i}{h}(P-E-z) u &= v+\frac{i}{h}[P,(\chi^b_i)^w] \tE(z)\,v\,dt\\
&= v + \chi_i^{b\prime} \tE(z)v - \sum_{j\in J_+(i)}\chi_j^{b\prime} \tE(z)v\,.
\end{align}
(we have used the splitting of the commutator explained above). 
Now, provided $T$ is not too large, the state $\tE(z)v$ is microlocalized away from $\Sigma_i$ so that $\chi_i^{b\prime} \tE(z)v=\cO(h^\infty)$
On the opposite, for each $j\in J_+(i)$ that state is a 
microlocal solution of $(P-E-z)u-0$ near $\tA_{ji}$, which can then be written as
$\cK_j(z)u_{-j}$ with $u_{-j}=R_{+j}\tE(z)v$. The above equality becomes
$$
\frac{i}{h}(P-E-z)\,u=v  - \sum_{j\in J_+(i)}R_{-j}u_{-j}\,,
$$
and solves \eqref{e:1'} microlocally.

If $v$ is microlocalized near $\Sigma_i$, we cutoff $\tE(z)v$ by $(\sum_{j\in J_-(i)}\chi^b_j+\chi_i^b)^w$, which is equivalent to identity 
near $\cT_i$, and take as above $u_{-j}=R_{+j}\tE(z)v$, $j\in J_+(i)$.

We have now fully inverted $\cP(z)$ microlocally near $K_E\times \prod_i\cT_i$, and the norm of the inverse can be shown to 
be of order unity. The effective Hamiltonian reads
$E_{-+}(z)=\cM(z)-Id$. Hence, as anticipated above, the existence of a nontrivial state $w$ satisfying $w=\cM(z)\,w$ is {\it equivalent} with
that of a microlocal solution to $(P-E-z)u=0$ near $K_E$.

To prove Thm~\ref{thm:main}, we must define our Grushin problem {\em globally}, that is properly define the auxiliary space
$\cH$ and the operators $R_{\pm}$, in such a way that $\cP(z)$ is invertible. One then says that the Grushin problem is {\em well-posed}.

\section{From the formal to the well-posed Grushin problem}\label{s:well-posed}

In order to make our Grushin problem well-posed, we will first ``deform'' the original Schr\"odinger operator $P(h)$ in order
to transform its resonances $z_i$ into {\em bona fide} $L^2$ eigenfunctions of the deformed operator $P_\theta(h)$. 
This deformation is performed through a 
``complex scaling'' of $P(h)$ far away from the scattering region. The operator $(P_\theta(h)-E)$ will now be elliptic 
outside a large ball $B(0,R)$. 
In order to enlarge this zone of
ellipticity to the complement of a smaller neighbourhood of $K_E$, we will then modify the 
Hilbert structure of our auxiliary states, using an appropriate {\it escape function} $G(x,\xi)$.
After these two modifications, we will be able to complete the construction of a well-posed Grushin problem, with
finite dimensional auxiliary spaces.

\subsection{Complex scaling}
Here we use the fact that outside a ball $B(0,R_0)\Supset \supp V$, one has $P(h)=-\frac{h^2}{2}\sum_{k=1}^n\frac{\partial^2}{\partial x_i^2}$. 
In that region that operator can 
be holomorphically exended into $\widetilde P=-\frac{h^2}{2}\sum_{k=1}^n\frac{\partial^2}{\partial z_i^2}$, acting on
functions on $\CC^n$. For $\theta>0$ small, we deform $\RR^n\subset\CC^n$ into a smooth contour $\Gamma_{\theta}\subset\CC^n$:
\begin{align*}
\Gamma_\theta \cap B_{\CC^n } ( 0 , R_0 ) &= B_{\RR^n } ( 0 , R_0 ) \,, \\
\Gamma_\theta \cap \CC^n \setminus B_{\CC^n } ( 0 , 2 R_0 ) &=
e^{ i \theta } \RR^n \cap \CC^n \setminus B_{\CC^n } ( 0 , 2 R_0 ) \,.
\end{align*}
We then define the operator $P_{\theta}(h)$ acting on $u \in \CIc ( \Gamma_{\theta } )$, by
$ P_{\theta} u  = \widetilde P ( \tilde u ) \rest_{\Gamma_{\theta} }$,
where $ \tilde u$ is an almost analytic extension of $ u $.
Through the identification
$\Gamma_\theta\ni x\longleftrightarrow (\sin\theta)^{-1}\Re x\in\RR^n$, the operator $P_\theta$ can considered as
acting on functions in $\CIc (\RR^n)$, with the action $-e^{-2i\theta}\frac{h^2\Delta}{2}$ outside $B(0,2R_0)$.
One can then show \cite{Sj} that the resolvent $ ( P_{\theta} - z )^{-1}:L^2\to H^2_h$ is meromorphic
in the region $\{\arg ( z ) > - 2 \theta\}$. The $L^2$ spectrum of $P_\theta(h)$ in that region
is discrete, independent of $ \theta $ and $ R$, and consists of the resonances of the initial operator $P(h)$.

Since $P_\theta(h)=P(h)$ inside the ball $B(0,R_0)\supset \pi K_E$, our formal Grushin problem remains unchanged
if we replace $P(h)$ by $P_\theta(h)$. 

Below we will take values of $\theta$ of the form $\theta\sim C\,h\log(1/h)$, $C>0$ fixed.

\subsection{Finite dimensional auxiliary spaces}\label{s:finite}

We have built in \S\ref{s:Grushin} a Grushin problem which is invertible microlocally near the
trapped set. To make that Grushin problem well-posed, we need
to make definite choices for the auxiliary spaces, that is for each $i=1,\ldots,J$ define
a subspace of $\cH_i\subset L^2(\RR^{n-1})$ containing the transversal data. 
This subspace should contain states microlocalized
in some neighbourhood $S_i$ of $\cT_i$, small enough to lie in the domain $\cup_{j\in J_+(i)}\tA_{ji}$ where $\kappa$ is defined. 
To construct this subspace explicitly, we may define the neighbourhood $S_i$ as $S_i=\{q_i(\rho)< 0\}$, for a well-chosen 
$q_i\in \CI(\RR^{n-1})$ satisfying $\liminf_{\rho\to\infty} q_i(\rho)>0$.  The subspace
$\cH_i$ can then be defined as the range of the spectral projector 
\be\label{e:projector}
\Pi_i\bbbone_{\RR_-}(q_i^w(y,hD_y))\,.
\ee 
According to Weyl's law,
for $h$ small enough the space $\cH_i$ has a finite dimension $\sim\vol(S_i)\,h^{-n+1}$.

One can then consider the Grushin problem \eqref{e:Grushin0}, with $P$ replaced by $P_\theta$,
the auxiliary space $\cH=\bigoplus_i\cH_i$ and the operators
\be\label{e:R_+}
R_{+i}(z)\defeq \Pi_i\cK_i^*(\bar z)\chi_i^{b\prime}\,,\qquad R_{-i}\defeq \chi_i^{b\prime}\cK_i(z)\Pi_i\,.
\ee
Unfortunately, when trying to solve this new Grushin problem, that is invert $\cP(z)$, one encouters difficulties.
Some of them are due to the fact that $\kappa$ does not leave the neighbourhoods $S_i$ invariant (see fig. \ref{fig:sections3-}). 
As a result, an initial 
datum $w_{i}\in\cH_i$ is propagated through $\cM_{ji}(z)$ into a state $\cM_{ji}(z)w_i$ which, in general, is not
microlocalized in $S_j$, and thus cannot belong to $\cH_j$. Brutally applying the projector $\Pi_j$
to $\cM_{ji}(z)w_i$ produces an extra term $(1-\Pi_j)\cM_{ji}w_i$, which is difficult to solve away.
Another difficulty arises when trying to solve the unhomogeneous problem \eqref{e:1'} for data $v$ microlocalized
at some distance from $K_E$.

\subsection{Escape functions and modified norms}\label{s:escape}
These difficulties can be tackled by modifying the Hilbert norms on $H^2_h(\RR^{n})$ and the auxiliary space $L^2(\RR^{n-1})^J$.
The new norms will be defined in terms of well-chosen {\it escape functions} $G\in\CIc(T^*\RR^n)$, $G^i\in\CIc(T^*\RR^{n-1})$.
Using these new norms, the problems mentioned above will disappear, because the states microlocalized 
away from $K_E$ will become easily solvable. 

Let us describe the escape function. For some small $\delta>0$, we consider the thickened energy shell
$\widehat{p^{-1}(E)}=\bigcup_{|s|\leq\delta} p^{-1}(E+s)$ and trapped set $\widehat K_E=\bigcup_{|s|\leq\delta} K_{E+s}$.
It is shown in \cite[\S\S 4.1,4.2,7.3]{SjZw04} and \cite[\S 6.1]{NZ2} that, for any small $\delta_0>0$ and large $R>0$,
and any neighbourhoods $U\subset \overline{U}\subset V$ of $\widehat K_E$, 
one can construct a function $G_0\in \CIc(T^*\RR^n)$ such that
\begin{align}
G_0= 0\quad \text{on $U$,}&\qquad
H_p G_0  \geq 0 \quad \text{on } T^*_{B(0,3R )}\RR^n, \\ 
\label{e:G_0} H_p G_0 \geq 1 \quad \text{ on }  T^*_{B(0,3R )}\RR^n \cap (\widehat{p^{-1}(E)} \setminus V),
&\qquad
H_p G_0   \geq -\delta_0 \quad \text{ on } T^* \RR^n\,. 
\end{align}
It is convenient\footnote{The role of this modification is to ultimately keep the norms 
$\|R_{+i}(z)\|_{H_{G}\to \cH_i}$, $\|R_{-i}(z)\|_{\cH_j\to H_{G}}$, $\|M_{ji}(z)\|_{\cH_i\to\cH_j}$ uniformly bounded} 
to slightly modify this function in the neighbourhood of the sets $S_i\subset\Sigma_i$.
Namely, we consider open neighbourhoods $\tilde W_i\Subset W_i$ of $S_i$ in $T^*\RR^n$, 
and modify $G_0$ into a function $G_1$, such that $H_pG_1=0$ in $\tilde W_i$ while $H_pG_1\geq 1$ on 
$T^*_{B(0,3R )}\RR^n \cap (\widehat{p^{-1}(E)} \setminus (V\cup\bigcup_iW_i)$.

We then set $G\defeq N h \log ( 1/ h ) G_1 $, with $N>0$ fixed but arbitrary large. 
The exponential $\exp ( G^w ( x, h D ) / h )$ is a pseudodifferential operator in some mildly exotic class, 
bounded and of bounded inverse on $L^2$, with norms $\cO(h^{-CN})$.
We call $H_G$ the vector space $H^2_h(\RR^n)$ equipped with the Hilbert norm
\be
\label{H_G}
\| u \|_{H_G } \defeq \| \exp ( - G^w(x,hD_x) /h ) u \|_{H^2_h}\,.
\ee
Similarly, we consider functions $G^i\in\CIc(T^*\RR^{n-1})$ such that (using the coordinate change $\tkappa_i$ near $\Sigma_i$)
$G^i(y',\eta')= G\circ\tkappa_i(0,y';0,\eta')$ in some neighbourhood of $S_i$, and modify the Hilbert norms 
on the space $L^2(\RR^{n-1})$ attached to the section $\Sigma_i$ by
\be
\|w_i\|_{H_{G^i}}\defeq \|e^{-(G^i)^w(y',hD_{y'})/h}w_i\|_{L^2(\RR^{n-1})}\,.
\ee

\subsection{How these norms resolve our problems}
Let us explain how this change of norm helps us.
The action of $P_\theta(h)$ on the Hilbert space $H_G$ is equivalent to the action of 
$P_{\theta,G}(h)\defeq e^{-G^w / h }\, P_\theta(h)\, e^{ G^w/h }$ on $H^2_h(\RR^n)$, which is a pseudodifferential operator with
symbol
$$
p_{\theta,G} (\rho)= p(\rho) - i N h\log(1/h) H_p G_1(\rho) +\cO(h^2 \log^2 ( 1/h ))\,,\quad \rho\in T^*_{B(0,R)}\RR^n\,.
$$
Provided we have chosen a dilation angle $\theta\ll  \delta_0 N h \log ( 1/h) $, the properties of $G_1$ show that
\be
\forall \rho\not\in (V\cup\bigcup_i W_i),\qquad 
| \Re p_{\theta,G }( \rho ) -E| \leq \delta/2 \Longrightarrow \Im p_{\theta,G} \leq - \theta/C\,,
\ee
This shows that, for any $z\in D(0,Ch)$, the symbol $(p_{\theta,G }-E-z)$ is invertible
outside $V\cup\bigcup_i W_i$, with inverse of order $(h\log h^{-1})^{-1}$. 
Hence, for any $v\in L^2$ microlocalized outside $V\cup\bigcup_i W_i$, the equation
$(P_{\theta,G}-E-z)u=v$ can be solved up to $\cO(h^\infty)$, with a solution $u$
microlocalized outside $V\cup\bigcup_i W_i$. 
This remark basically tackles the second problem mentioned at the end of \S\ref{s:finite}.

The first problem (the fact that $\kappa_{ji}(S_i)$ is not contained in $S_j$) 
is also resolved through this change of norms. Indeed, 
the escape function $G_1$ can be chosen such that it {\it uniformly increases} (say, by some $2C>0$)
along all trajectories of the form
$\rho\in S_i\mapsto \kappa_{ji}(\rho)\in\Sigma_j\setminus\in S_j$, so that  $\frac{e^{-G(\kappa_{ji}(\rho))/h}}{e^{-G(\rho)/h}}\leq h^{2NC}$.
This implies that, for any state $w_i$ microlocalized
near such a point $\rho$, one gets (taking the definition \eqref{e:projection-i} for $R_{+j}$)
\be\label{e:damping}
\|R_{+j}(z)\cK_{i}(z)w_i\|_{H_{G^j}}=\cO(h^{NC})\,\|w_i\|_{H_{G^i}}\,.
\ee
We then need to modify the finite rank projectors \eqref{e:projector} defining our auxiliary states, such as to make
them orthogonal w.r.to the new norms (otherwise $\|\Pi_i\|_{H_{G^i}\to H_{G^i}}$ could be very large). 
This modification only amounts to adding a subprincipal (complex valued)
term to the function $q_i$, such that $q_i^w$ becomes selfadjoint on $H_{G^i}$, and the spectral
projector \eqref{e:projector} orthogonal.
The space $\cH_i\defeq \Pi_i H_{G^i}$ is still made of states microlocalized in $S_i$, and has dimension $\sim\vol(S_i) h^{-n+1}$.
Our operators $R_{\pm i}$ will be defined by \eqref{e:R_+}. 

Let us reconsider the homogeneous problem \S\ref{s:homog} with data $w_i\in\cH_i$ in our new Grushin problem.
For $j\in J_+(i)$, the state $\cM_{ji}(z)\,w_i$ does not a priori belong to $\cH_j$. However,, the estimate \eqref{e:damping} shows
that the component of $\cM_{ji}(z)\,w_i$ microlocalized outside $S_j$ has an $H_{G^j}$-norm of order $\cO(h^{NC})$.
As a result, defining the finite rank operators 
\be\label{M_ij}
M_{ji}(z)\defeq \Pi_j\,\cM_{ji}(z):\cH_i\to\cH_j\,,
\ee
we find that
$$
u_{-i}\defeq -w_i + \sum_{j\in J_+(i)}M_{ij}(z)w_j\in\cH_i
$$ 
provides a solution to the homogeneous problem, up to an error $\cO(h^{NC})(\sum_i\| w_i\|_{\cH_i})$.

The nonhomogeneous problem \eqref{e:1'} can be solved as well, up to a comparable error (see \cite{NSZ1} for details).

To summarize, our globally defined Grushin problem has an
approximate inverse $\cE(z)$:
$$
\cP(z)\cE(z)=I+\cR(z),\qquad \|\cR(z)\|_{L^2\times \cH\to L^2\times \cH}=\cO(h^{NC})\,,
$$
where we insist on the fact that $N$ can be chosen arbitrary large (it comes from the factor in front of the escape function $G$). 
Hence, for $h$ small enough this operator 
has the exact inverse 
$\tilde \cE(z)=\cE(z) (I+\cR(z))^{-1}=\cE(z)+\cO_{L^2\times\cH\to H^2_h\times\cH}(h^{NC})$. In particular, the
lower-right entry of $\tilde\cE(z)$ (that is, the exact effective Hamiltonian) reads
$$
\tE_{-+}(z)=I-M(z)+\cO_{\cH\to \cH}(h^{NC})\,,
$$
where $M(z)$ is the matrix composed of the finite dimensional operators \eqref{M_ij}.

As explained in \S\ref{e:Grushin-general}, this exact inversion implies that the eigenvalues $\{z_i\}$ of $(P_\theta-E)$
in $D(0,Ch)$ coincide (with multiplicities) with the zeros of $\det(E_{-+}(z))$.
\hfill$\square$

%%%%%%%%%%%%%%%%%%%%%%%%%%%%%%%%%%
\medskip

\noindent
{\sc Acknowledgements.} The author has been partially supported by the Agence Nationale de la Recherche through the
grant ANR-09-JCJC-0099-01. These notes were written while the author was visiting the Institute of Advanced Study in Princeton,
supported by the National Science Foundation under agreement No. DMS-0635607.

%%%%%%%%%%%%%%%%%%%%%%%%%%%%%%%%%%

\end{document}